\documentclass[showpacs,amsmath,amssymb,twocolumn,floatfix,pra]{revtex4}
\usepackage[dvips]{graphicx}
\begin{document}
\title{ Entanglement versus relaxation and decoherence\\
  in a quantum algorithm for quantum chaos}
\author{S. Bettelli, and D. L. Shepelyansky}    
\homepage{http://www.quantware.ups-tlse.fr}
\affiliation{Laboratoire de Physique Quantique, UMR 5626 du CNRS,
  Universit\'e Paul Sabatier, 31062 Toulouse Cedex 4, France}
\date{\today}

\begin{abstract}
  We study analytically and numerically the behavior of the concurrence
  (a measure of the entanglement of formation) of a pair of qubits in a
  quantum computer operating an efficient algorithm for quantum chaos.
  Our results show that in an ideal algorithm the entanglement decays
  exponentially with the diffusive relaxation rate induced by classical
  chaos. This decay reaches a residual level which drops exponentially
  with the number of qubits $n_q$. Decoherence destroys the residual
  entanglement with a rate exponential in $n_q$.
\end{abstract} 
\pacs{03.67.Lx, 05.45.Mt}  

\maketitle

Enormous interest into quantum information and computation
\cite{ChuangNielsen00} has generated serious efforts in characterizing and
understanding quantum entanglement which is considered as the ultimate
origin of quantum power (see a recent review \cite{Bruss02}). A quantitative
measure of the entanglement of formation, namely the concurrence $C$, was
introduced and shown to be able to characterize an arbitrary state of two
qubits \cite{HillWootters97,Wootters98}. 
Being closely related to the von Neumann entropy $S$ of the reduced density
matrix $\rho$ of two qubits%
\footnote{Given a density matrix $\rho$ for a pair of two-level systems, the
  concurrence $C$ is defined as $min\{0,\lambda_1-\lambda_2-\lambda_3
  -\lambda_4\}$, where the $\lambda_i$ are the eigenvalues, in decreasing
  value order, of the hermitian matrix $\sqrt{\sqrt{\rho}\widetilde{\rho}
    \sqrt{\rho}}$, with $\widetilde{\rho}=(\sigma_y\otimes\sigma_y)\rho^*
  (\sigma_y\otimes\sigma_y)$. The von Neumann entropy $S$ of $\rho$ 
  monotonically increases from 0 to 1 when $C$ goes from 0 to 1
  \cite{HillWootters97,Wootters98}.},
this quantity was recently found to have interesting applications to quantum
phase transitions in interacting spin systems \cite{FazioEtal02}. In parallel,
the properties of entanglement were investigated in a quantum model of coupled
tops where it was shown that 
there exists a typical value of entanglement which is determined by the
chaotic behavior of the dynamics of the model \cite{MillerSarkar99} and
that the growth rate of entanglement of initially decoupled tops is increased
by the underlying classical chaos \cite{BandyLaksh02}.
On the same line, it was recently shown that contrary to intuition even
a heat bath may create entanglement between two qubits \cite{Braun02}.

All these studies \cite{FazioEtal02,MillerSarkar99,BandyLaksh02,Braun02}
clearly demonstrated how rich entanglement properties can be in interacting
quantum systems. However, in the context of quantum computation it is much
more crucial to analyze the evolution of entanglement in a specific algorithm
performing an operational task. Indeed, it is expected that the entanglement
is very sensitive to noise and decoherence \cite{PazZurek01,SchackCaves93,%
SchackCaves96} and the understanding of its behavior in an operating algorithm
can lead to better strategies in the control of decoherence and imperfection
effects. As for our knowledge such direct investigations have not been
performed until now. Therefore, in this paper we study the behavior of the
concurrence in an efficient algorithm for the quantum sawtooth map
which has been proposed recently in \cite{BenentiEtal01}.
The algorithm for this model has a number of important advantages: all $n_q$
qubits are used in an optimal way and no ancillae are required, one map
iteration in the Hilbert space of size $N=2^{n_q}$ is performed in $O(n_q^2)$
quantum gates, and the algorithm is based on the quantum Fourier transform
(QFT) which is one of the main elements of various quantum algorithms
\cite{ChuangNielsen00}. This allows to simulate a complex dynamics in the
regime of quantum chaos with a small number of qubits.
Since the entanglement can be efficiently measured experimentally (see e.g.
\cite{HorodeckiEkert02}) the experimental observation of the concurrence
behavior discussed here can be realized on NMR \cite{CoryEtal01,ChuangEtal01}
or ion-trap \cite{BlattEtal03} based quantum computers with about 6 - 10
qubits and a few hundreds of gates.

Contrary to the situation discussed in \cite{MillerSarkar99,BandyLaksh02},
our results show that in the exact quantum algorithm the underlying classical
chaos leads to an exponential decrease of the concurrence $C$ down to some
residual level $\bar{C}$ which characterizes the global system coherence.
On the other hand, the presence of noise in the quantum gates leads to a
destruction of this coherence with a rate $\Gamma$ growing exponentially
with the number of qubits $n_q$. This shows that entanglement can be very
sensitive to decoherence.

The dynamics of the classical sawtooth map \cite{DanaEtal89,DanaEtal90}
is given by:
\begin{equation}
  \overline{n} = {n} - k\frac{dV(\theta)}{d\theta},
  \quad
  \overline{\theta} = \theta + T\overline{n} \mod 2\pi
  \label{classical-sawtooth}
\end{equation}
where $V(\theta)=-\theta^2/2$, $-pi\leq\theta<\pi$ and the bars denote the
variables after one iteration. After rescaling $y=Tn$ and $x=\theta$, it
is clear that the dynamics depends only on the parameter $K=kT$. Due to the
discontinuity in the derivative of $V(\theta)$, the Kolmogorov-Arnold-Moser
(KAM) theorem cannot be applied to the map (\ref{classical-sawtooth}) and
its dynamics becomes chaotic and diffusive for arbitrarily small values of
the chaos parameter $K>0$ \cite{DanaEtal89}. For $K \ll 1$ the diffusion is
governed by a nontrivial cantori regime which has been worked out in
\cite{DanaEtal89}. In this case the rescaled diffusion rate $D_0(K)=
(\Delta y)^2/t \approx 1.2 \pi^2 K^{2.5}/3$ is much smaller than the
quasi-linear diffusion rate corresponding to the random phase approximation
$D_{ql} =\pi^2 K^2/3$ (the latter becomes valid only at $K \gg 1$).
The diffusion rate in $n$ is $D=D_0(K)/T^2$.

The quantum sawtooth map \cite{BenentiEtal01,BorgonoviEtal96,Borgonovi98,%
  CasatiProsen99,PrangeEtal99} is the quantized version of the classical
map, to which it corresponds in the limit $k\to \infty$, $T\to 0$, and
$K=kT=\hbox{const}$. One step in the quantum map is given by the unitary
operator $\hat{U}$ acting on the wave function $\psi(\theta)$:
\begin{equation}
  \overline{\psi(\theta)} = \hat{U}\psi(\theta) =
  e^{ik{\theta}^2/2} e^{-iT\hat{n}^2/2} \psi(\theta),  
  \label{quantum-sawtooth}
\end{equation}
where $\hat{n} = -i\partial/\partial\theta$ (we set $\hbar=1$);
when $n_q$ qubits are used for the discretization, so that the total
number of levels is $N=2^{n_q}$, we assume periodic boundary conditions
on a torus in the phase representation ($\psi(\theta+2\pi)=\psi(\theta)$)
and in the momentum representation ($\psi(n+N)=\psi(n)$), as discussed
in \cite{BenentiEtal01,Izrailev90}. As a result, the phase $\theta$ takes
only $N=2^{n_q}$ discrete equidistant values in the interval $-\pi\leq
\theta<\pi$, and so does the momentum ($n=0,1,...,N-1$, i.e. $n+N$ is
identified with $n$).
The sawtooth map approximately describes the dynamics in the stadium
billiard \cite{BorgonoviEtal96}, the phenomenon of dynamical localization
which is similar to the Anderson localization in disordered potentials and
cantori induced localization \cite{Borgonovi98,CasatiProsen99,PrangeEtal99}.
Thus this simple map describes a rich and complex dynamics and 
represents an interesting testing ground for efficient quantum computation.
It is especially important to understand how this complex quantum dynamics
will be affected by imperfections in realistic quantum computers.

The numerical simulation of the map (\ref{quantum-sawtooth}) is based on
the quantum algorithm described in \cite{BenentiEtal01} and is implemented on
the basis of a quantum computer language developed in \cite{BettelliEtal03}
which is well adapted for the experimental operation of quantum gates
via classical computer software. In this way the dynamics of up to 20
qubits can be easily simulated on a laptop.

\begin{figure}
  \includegraphics[width=\linewidth]{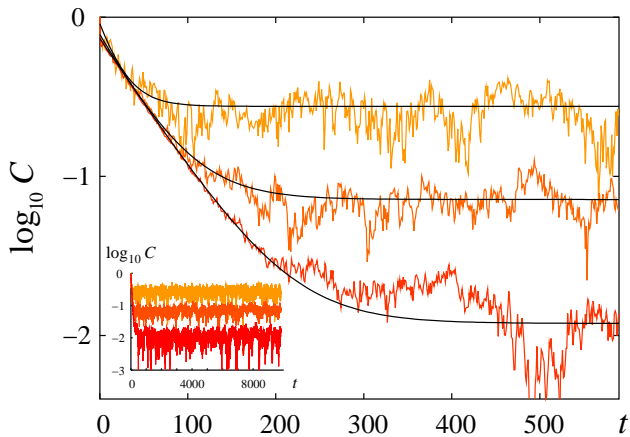}
  \caption{Dependence of the concurrence $C$ on the dimensionless time $t$
    (the number of map iterations) for the map (\ref{quantum-sawtooth}) at
    $K=0.5, L=4$ and $n_q=8,12,16$ (curves from top to bottom respectively).
    The smooth curves show the fit $C(t)=A\exp(-\gamma t)+{\bar C}$ of the
    relaxation to the asymptotic value ${\bar C}$ obtained on a larger time
    interval ($t \leq 10^4$).
    The inset shows $C(t)$ on larger time scale. The initial state is
    $(|00\rangle + |11\rangle)|\phi\rangle/\sqrt{2}$ where $|\phi\rangle$
    is the uniform superposition of all but the two most significant
    qubits. Here and below the logarithms are decimal and all axis units
    are dimensionless.
  }
  \label{fig1}
\end{figure} 

To investigate the behavior of the concurrence in the quantum map
(\ref{quantum-sawtooth}) we compute $C$ for the two most significant
qubits which determine the first two binary digits $a_{1,2}$ in the
expansion of momentum $n$: the reduced density matrix $\rho$ for this
qubit pair is obtained by tracing out  all other $n_q-2$ less significant
qubits (the digits $a_i$ with $3 \leq i \leq n_q$ in the expansion of
$n=(a_1 a_2 a_3 ...a_i... a_{n_q})$). After that $C$ is computed from
$\rho$ as described in \cite{HillWootters97,Wootters98}.
In this way we obtain the concurrence value $C$ on a global scale of the
whole system which is decomposed into 4 equal parts with $N/4$ quantum
states in each of them. In addition we fix $T=2\pi L/N$ in the regime of
quantum resonance so that $L$ gives the integer number of classical phase
space cells  embedded in the quantum torus of size $N$
\cite{BenentiEtal01,Izrailev90} (the classical dynamics is periodic in
$n$ with period $2\pi/T$). In the following we also take $L$ to be a
multiple of 4 to have an integer number of classical cells in the 4
parts of the partition in the momentum $n$.

Typical examples of the dependence of $C$ on the number of map iterations
$t$ are shown in Fig.1. According to these data $C(t)$ decays exponentially
down to a residual value $\bar{C}$ and, in the limit of large $N$, the decay
rate $\gamma$ becomes independent of $N$. It is natural to compare this rate
with the rate of classical relaxation. Indeed, due to underlying classical
chaos, the probability distribution $f_n=|\psi_n|^2$ over $n$ is described
by the Fokker-Planck equation $\partial f_n/\partial t = D \partial^2 f_n/
\partial^2 n /2$ which gives the relaxation to equipartition with the rate
\begin{equation}
  \gamma_c = 2 \pi^2 D/N^2 = D_0(K)/2L^2 \;\; .
  \label{eq3}
\end{equation}
The comparison between this  classical value $\gamma_c$ and the rate of
concurrence decay $\gamma$ is given in Fig.2. It clearly shows that the
decay rate of $C(t)$ is given by the classical rate: $\gamma=\gamma_c$.
It is important to stress that this relation remains valid also in the
nontrivial cantori regime ($K \ll 1$) and that the quantum decay reproduces
all oscillations of the classical diffusion (see inset in Fig.2).

\begin{figure}
  \includegraphics[width=\linewidth]{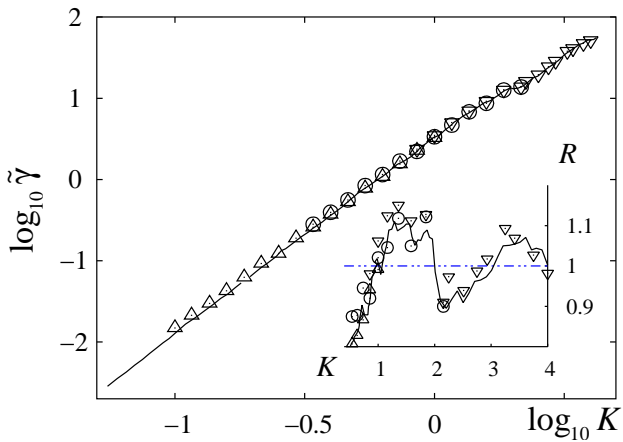}
  \caption{Dependence of the rescaled rate of the concurrence decay
    $\tilde \gamma = 2 \gamma L^2$ on the chaos parameter $K$ for $n_q=19,
    L=16$ (triangles down); $n_q=18, L=8$ (circles) and $n_q=17, L=4$
    (triangles up). The solid curve gives the values of the diffusion
    rate $D_0(K)$ taken from Figs.2,3 of \cite{DanaEtal89} showing that
    $\gamma$ is determined by the classical relaxation rate (\ref{eq3}).
    The inset shows data on a larger scale with $R=\tilde \gamma /D_{ql}$
    (symbols) and $R=D_0(K) /D_{ql}$ (curve from \cite{DanaEtal89}).
  }
  \label{fig2}
\end{figure} 

The properties of the residual value of the concurrence $\bar C$ are analyzed
in Fig.3. We will argue in the following that they can be understood in terms
of the system conductance $g$ (\cite{EdwardsThouless72, Thouless74});
in view of this we express $\bar C$ vs. $g = 2\gamma_c/\Delta = ND_0(K)/L^2$
where, up to a constant factor, $\Delta=1/N$ is the level spacing and
$2\gamma_c=D_0(K)/L^2$ is the Thouless energy (see \cite{Mirlin00} for
a recent review).
In spite of strong fluctuations the data presented in Fig.3 can be described
by  the global average dependence $\bar C \sim 1/\sqrt{g} \propto 1/\sqrt{N}$.
Indeed, for $K=0.5, L=4$ the system size varies by 3 orders of magnitude and
the fit gives an algebraic decay with power $\alpha = 0.56\pm0.02$ being
close to $1/2$. We attribute the presence of strong fluctuations to the fact
that the value $\bar C$ is averaged only over time but there is no averaging
over parameters. Thus, from the point of view of disordered systems
\cite{Mirlin00} $\bar C$ represents only one value for one realization
of disorder.

\begin{figure}
  \includegraphics[width=\linewidth]{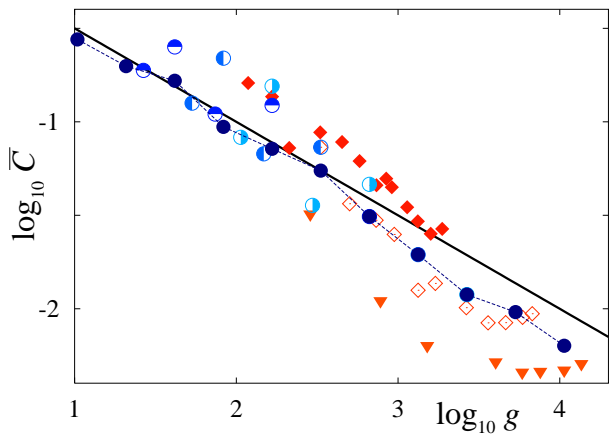}
  \caption{Dependence of the residual value of the concurrence $\bar C$
    on the system conductance $g = N D_0(K)/L^2$ for a broad range of
    parameters: half filled circles show dependence on $L=4, 8, 12, 16, 20$
    for $K=0.5$ and $n_q=14, 15, 16$; diamonds and triangles show the
    variation with $K$ for $n_q=14, L=16$; $n_q=15, L=8$ and $n_q=16, L=4$.
    The filled circles connected by the dashed curve show the dependence 
    on $N$ for $K=0.5, L=4$. The solid line marks the slope $1/\sqrt{g}$.
  }
  \label{fig3}
\end{figure} 

We propose the following explanation of the results presented in Figs.1-3.
For a state $|\psi\rangle$ like in Fig.1, we can write  $|\psi\rangle =
\sum_{a_1a_2}|a_1a_2\rangle |\phi_{a_1a_2}\rangle$ where $a_{1,2}=0$ or $1$.
Then the value of the concurrence $C$ is proportional to the difference of
two scalar products, $C \sim |Q_{14} - Q_{23}|$, where
$Q_{14}=2\sqrt{|\langle\phi_{00}|\phi_{11}\rangle|^2}$ 
and $Q_{23}=2\sqrt{|\langle\phi_{01}|\phi_{10}\rangle|^2}$.
From this relation and the fact that the initial state is symmetrically
distributed with respect to the transformation $n \to N - n$, it follows
that $C$ is proportional to the difference $|W_{11}+W_{00}-W_{01}-W_{10}|$
where $W_{a_1a_2}$ is the total probability inside the part $(a_1a_2)$.
In the classical limit this probability difference relaxes to zero with the
classical relaxation rate $\gamma_c$, and that's why $\gamma = \gamma_c$ in
agreement with the data of Fig.2.

The residual value $\bar{C}$ is determined by the quantum fluctuations of the
previous difference of scalar products. In fact, due to the discretization of
map (\ref{quantum-sawtooth}), the symmetry $n \to N - n$ is broken and
$|\phi_{00}\rangle$ becomes different from $|\phi_{11}\rangle$.
Therefore in the scalar product $Q_{14}$ (and $Q_{23}$) the $N/4$ terms have
random signs and thus $Q_{14} \propto 1/\sqrt{N}$ (each term is of the order
of $1/N$). In this estimate we assumed a summation over all $N$ wave function
components. However for finite values of the conductance $g$ only the states
inside the Thouless energy interval $2 \gamma_c$ have a significant scalar
overlap \cite{Mirlin00,Blanter96}, and thus we can make a conjecture that
$N$ should be replaced by the effective number of components which is of the
order of $N_{\textrm{eff}} \sim \gamma_c/\Delta \sim g$. According to this,
$\bar{C} \sim 1/\sqrt{g}$ in agreement with the data of Fig.3. 

The existence of a residual level of concurrence for an ideal quantum
algorithm reflects the fact that the global behavior of the whole system
remains coherent. In fact, the Poincar\'e theorem guarantees that for very
large times the concurrence will have a revival close to the initial value
(however, this will happen on an exponentially large time scale).
The situation becomes qualitatively different in presence of external
decoherence represented by noisy gates. In our numerical simulations,
noisy gates are modeled by unitary rotations by an angle randomly
fluctuating in the interval $(-\epsilon/2, \epsilon/2)$ around the perfect
rotation angle. 
The presence of this external decoherence leads to a decrease of the residual
value of $C$ as illustrated in Fig.4: the constant level is replaced by an
exponential decay which gives $\bar{C} \propto \exp(-\Gamma t)$. 

\begin{figure}
  \includegraphics[width=\linewidth]{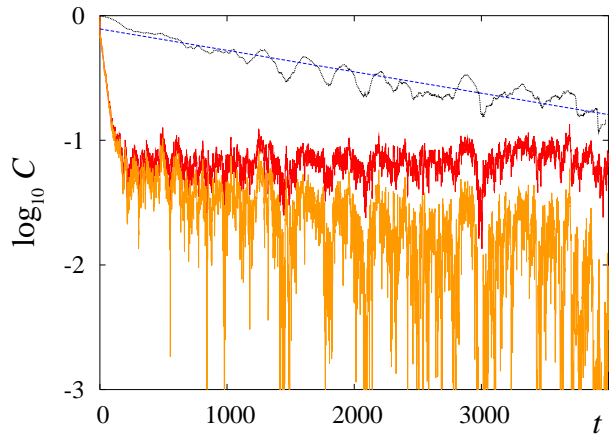}
  \caption{Effects of decoherence on the residual concurrence: the two
    lower curves show the dependence $C(t)$ for an ideal algorithm (top)
    and an algorithm with noisy gates at noise amplitude $\epsilon =0.003$
    (bottom). The time $t$ is dimensionless (it is the number of map steps).
    In the latter case the average is done over 20 noise realizations. The
    curve in the upper part shows the ratio of $C(t)$ at $\epsilon=0.003$
    to its value in an ideal algorithm, this ratio is averaged over a
    100-kick moving window to reduce fluctuations. The dashed straight
    line shows a fit of the ratio to an exponential decay proportional
    to $e^{- \Gamma t}$. Here $n_q=12, K=0.5, L=4$.
  }
  \label{fig4}
\end{figure} 

In order to obtain the dependence of $\Gamma$ on the parameters we extracted
it from the fit of the averaged ratio of $C$ under a noisy evolution to its
value in the ideal one. An example of such ratio and the corresponding fit is
shown in Fig.4. To suppress fluctuations we averaged over $20$ realizations
of the noisy evolution. Moreover, the fit was restricted to the plateau
regime, where the exact concurrence is fluctuating around its residual value
(the initial diffusive relaxation was excluded from the fit). 

\begin{figure}
  \includegraphics[width=\linewidth]{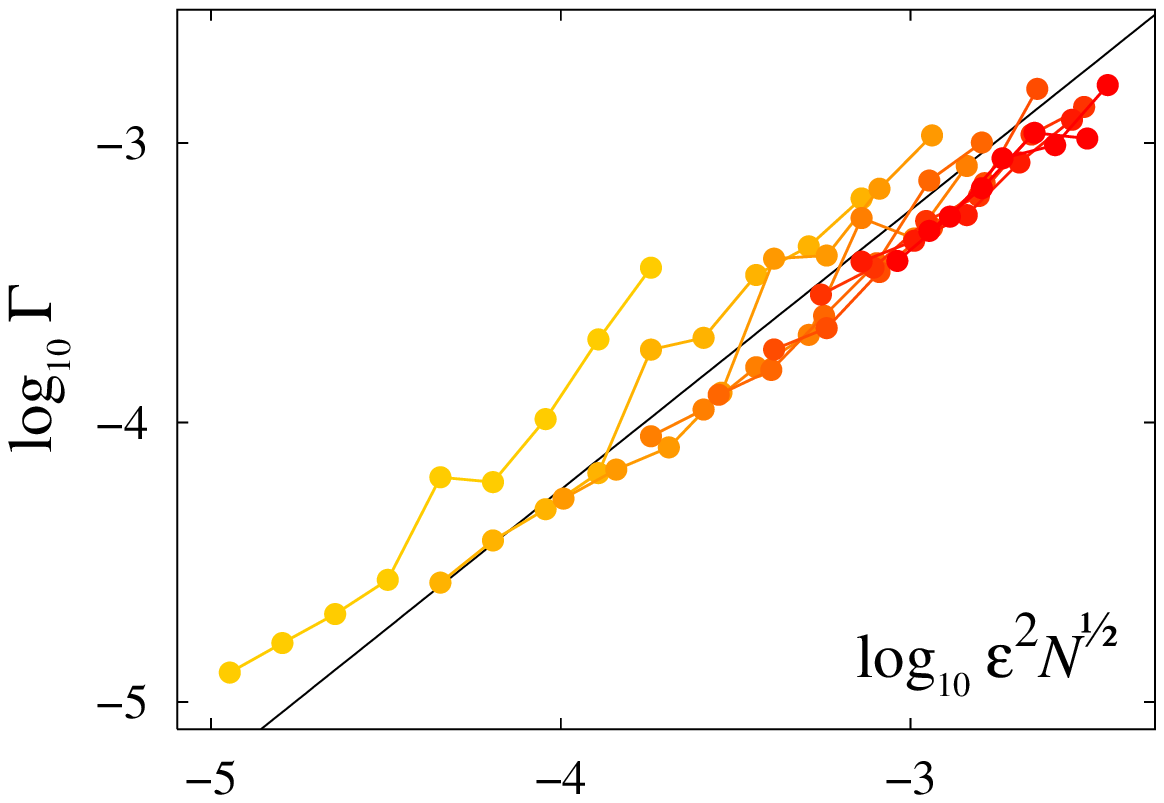}
  \caption{Dependence of the decoherence induced decay rate $\Gamma$ of
    the residual concurrence on $\epsilon^2\sqrt{N}$ for $K=0.5$, $L=4$.
    Here the noise amplitude $\epsilon$ changes from $0.001$ to $0.01$
    (10 equidistant values) for $7 \leq n_q \leq 15$. The data points
    (circles) are connected by lines for fixed values of $\epsilon$.
    The color intensity  changes gradually from one chain to another to
    mark the variation of $\epsilon$ (low/high intensity corresponds to
    small/large values of $\epsilon$). The straight line shows the averaged
    behavior $\Gamma = 0.58\epsilon^2\sqrt{N}$.}
  \label{fig5}
\end{figure} 

The results for $\Gamma$ obtained in this way are presented in Fig.5. Quite
naturally we find that $\Gamma \propto \epsilon^2$, as it was also seen in
other simulations of quantum algorithms with noisy gates (e.g.
\cite{ChepShep02}).This scaling becomes better and better for large
$\epsilon$ values where $\Gamma$ is larger. However, more surprisingly
there is an exponential growth of $\Gamma$ with the number of qubits $n_q$
($\Gamma \propto \sqrt{N}$). This result is very different from those
obtained in \cite{BenentiEtal01,ChepShep02}, where the time scale for
fidelity and the decoherence rate for tunneling oscillations varied
polynomially with $n_q$. We see two possible reasons for the exponential
sensitivity of the residual concurrence to decoherence.
At first, in our case $\Gamma$ is computed over a very large time interval,
for which the quantum dynamics already reached its asymptotic behavior
(plateau for the residual concurrence); it is known that on very large times
the eigenstates are exponentially sensitive to imperfections due to the
chaotic structure of the wave functions (see results and discussions in
\cite{BenentiEtal02}). Another possible reason can be related to the fact
that the residual value of the concurrence on the plateau is on its own
exponentially small and maybe this is the reason why it becomes so sensitive
to decoherence. Further investigation on the decoherence effects for the
concurrence are required to understand in a better way this exponential
sensitivity of $C$.

In our studies we restricted ourselves to the investigation of entanglement
only between two qubits. The problem of the characterizing the entanglement
of a larger number of qubits represents an interesting challenge but at
the same time it is much more complicated \cite{Bruss02}. However even the
relatively simple case of two qubits shows nontrivial links between
concurrence and such interesting physical phenomena as quantum phase
transitions \cite{FazioEtal02} and statistical relaxation.

In summary, our studies show that the decay of the concurrence in an
operating quantum computer is determined by the underlying relaxation rate
of the classical dynamics. We show that the residual level of entanglement
in an ideal algorithm scales as the inverse square root of the conductance
of the system. This residual entanglement is destroyed by decoherence, whose
effective rate grows exponentially with the number of qubits.

This work was supported in part by the EC contracts RTN  QTRANS and
IST-FET EDIQIP and the NSA and ARDA under ARO contract No. DAAD19-01-1-0553.

\end{document}